\documentclass[12pt,reqno]{amsart}
\usepackage{graphicx}
\usepackage{amscd}
\usepackage{amsmath}
\usepackage{epsfig}
\usepackage{amsfonts}
\usepackage{amssymb}

\providecommand{\U}[1]{\protect\rule{.1in}{.1in}}
\providecommand{\U}[1]{\protect\rule{.1in}{.1in}}
\allowdisplaybreaks

\theoremstyle{plain}

\numberwithin{equation}{section}

\input{tcilatex}

\begin{document}
\title[The Degenerate Parametric Oscillator ]{The Degenerate Parametric
Oscillator and Ince's Equation}
\author{Ricardo Cordero-Soto}
\address{Mathematical, Computational and Modeling Sciences Center, Arizona
State University, Tempe, AZ 85287--1904, U.S.A.}
\email{ricardojavier81@gmail.com}
\author{Sergei K. Suslov}
\address{Mathematical, Computational and Modeling Sciences Center, Arizona
State University, Tempe, AZ 85287--1904, U.S.A.}
\email{sks@asu.edu}
\urladdr{http://hahn.la.asu.edu/\symbol{126}suslov/index.html}
\date{\today }

\begin{abstract}
We construct Green's function for the quantum degenerate parametric
oscillator in terms of standard solutions of Ince's equation in a framework
of a general approach to harmonic oscillators. Exact time-dependent wave
functions and their connections with dynamical invariants and $SU\left(
1,1\right) $ group are also discussed.
\end{abstract}

\maketitle

\section{The Degenerate Parametric Amplifier Hamiltonian}

We use standard annihilation and creation operators in the coordinate
representation \cite{Flu}, \cite{La:Lif}, \cite{Merz} given by%
\begin{equation}
\widehat{a}=\sqrt{\frac{m}{2\hslash \omega }}\left( \omega x+\frac{ip}{m}%
\right) ,\qquad \widehat{a}^{\dagger }=\sqrt{\frac{m}{2\hslash \omega }}%
\left( \omega x-\frac{ip}{m}\right)   \label{at1}
\end{equation}%
with $p=-i\hslash \partial /\partial x$ and $\widehat{a}\widehat{a}^{\dagger
}-\widehat{a}^{\dagger }\widehat{a}=1$ throughout this Letter. The following
Hamiltonian%
\begin{equation}
H\left( t\right) =\frac{\hslash \omega }{2}\left( \widehat{a}\ \widehat{a}%
^{\dagger }+\widehat{a}^{\dagger }\ \widehat{a}\right) -\frac{\hslash
\lambda }{2}\left( e^{2i\omega t}\ \widehat{a}^{2}+e^{-2i\omega t}\left( 
\widehat{a}^{\dagger }\right) ^{2}\right)   \label{at2}
\end{equation}%
of the time-dependent Schr\"{o}dinger equation%
\begin{equation}
i\hslash \frac{\partial \psi }{\partial t}=H\left( t\right) \psi 
\label{at2a}
\end{equation}%
describes the process of \ degenerate parametric amplification in quantum
optics. The first term corresponds to the self-energy of the oscillator
representing the mode of interest, the second term describes the coupling of
the classical pump to that mode (with the phenomenological constant $\lambda
),$ giving rise to the parametric amplification process (the phase of the
pump is taken to be zero at $t=0$ for convenience). Takahasi \cite{Taka65}
wrote a detailed paper on this subject as far as in 1965 (see also \cite%
{Louisel64}, \cite{Orszag}, \cite{Per:Hra:Jur}, \cite{Puri}, \cite{Raiford70}%
, \cite{Raiford74}, \cite{Shen}, \cite{Stoler74}, \cite{Walls:Milburn} and
references therein). The same Hamiltonian had also been considered by
Angelow and Trifonov in the mid 1990s in order to describe the light
propagation in the nonlinear $Ti:LiNbO_{3}$ anisotropic waveguide (details
can be found in \cite{Angelow98} and \cite{Angelow:Trifonov95}). Another
form of this variable quadratic Hamiltonian is given by%
\begin{eqnarray}
H &=&\frac{1}{2m}\left( 1+\frac{\lambda }{\omega }\cos \left( 2\omega
t\right) \right) p^{2}+\frac{m\omega ^{2}}{2}\left( 1-\frac{\lambda }{\omega 
}\cos \left( 2\omega t\right) \right) x^{2}  \label{at3} \\
&&+\frac{\lambda }{2}\sin \left( 2\omega t\right) \left( px+xp\right)
,\qquad p=\frac{\hslash }{i}\frac{\partial }{\partial x}.  \notag
\end{eqnarray}%
We refer to (\ref{at2})--(\ref{at2a}) and/or (\ref{at3}) as the degenerate
parametric oscillator \cite{Orszag}. It can be recognized as a special case
of so-called generalized harmonic oscillators that had attracted
considerable attention over the years in view of their great importance to
many advanced quantum problems including coherent states and uncertainty
relations, Berry's phase, asymptotic and numerical methods, quantization of
mechanical systems, Hamiltonian cosmology, charged particle traps and motion
in uniform magnetic fields, molecular spectroscopy and polyatomic molecules
in varying external fields. Quadratic Hamiltonians have particular
applications in quantum electrodynamics because the electromagnetic field
can be represented as a set of forced harmonic oscillators. See, for
example, \cite{Berry85}, \cite{Cor-Sot:Sua:SusInv}, \cite{Dod:Mal:Man75}, 
\cite{Dodonov:Man'koFIAN87}, \cite{FeynmanPhD}, \cite{Fey:Hib}, \cite%
{GomezVergel:Villasenor09}, \cite{Hannay85}, \cite{Leach90}, \cite%
{Malkin:Man'ko79}, \cite{Menouar:Maamache:Choi10}, \cite{Suslov10}, \cite%
{Wolf81}, \cite{Yeon:Lee:Um:George:Pandey93} and references therein.
Nonlinear oscillators play a central role in the theory of Bose--Einstein
condensation \cite{Dal:Giorg:Pitaevski:Str99} because the dynamics of gases
of cooled atoms in a magnetic trap at very low temperatures can be described
by an effective equation for the condensate wave function known as the
Gross--Pitaevskii (or nonlinear Schr\"{o}dinger) equation \cite%
{Kagan:Surkov:Shlyap96}, \cite{Kagan:Surkov:Shlyap97}, \cite%
{Kivsh:Alex:Tur01}, \cite{Per-G:Tor:Mont} and \cite%
{Trall-Gin:Drake:Lop-Rich:Trall-Herr:Bir}.\medskip 

A goal of this Letter is to construct Green's function of the Hamiltonian (%
\ref{at3}) in the coordinate representation in terms of solutions of Ince's
equation studied in \cite{Mag:Win} and \cite{Menn68} (see also references
therein). Exact time-dependent wave functions and connections with linear
and quadratic dynamical invariants and $SU\left( 1,1\right) $ group are also
briefly discussed.

\section{Generalized Harmonic Oscillators}

The degenerate parametric oscillator Hamiltonian (\ref{at3}) belongs to
quantum systems described by the one-dimensional time-dependent Schr\"{o}%
dinger equation%
\begin{equation}
i\frac{\partial \psi }{\partial t}=H\psi  \label{Schroedinger}
\end{equation}%
with variable quadratic Hamiltonians of the form%
\begin{equation}
H=a\left( t\right) p^{2}+b\left( t\right) x^{2}+c\left( t\right) px+d\left(
t\right) xp,\quad p=-i\frac{\partial }{\partial x},  \label{GenHam}
\end{equation}%
where $a\left( t\right) ,$ $b\left( t\right) ,$ $c\left( t\right) ,$ and $%
d\left( t\right) $ are real-valued functions of time, $t,$ only (see, for
example, \cite{Cor-Sot:Lop:Sua:Sus}, \cite{Cor-Sot:Sua:Sus}, \cite%
{Cor-Sot:Sua:SusInv}, \cite{Cor-Sot:Sus}, \cite{Dod:Mal:Man75}, \cite%
{FeynmanPhD}, \cite{Feynman}, \cite{Fey:Hib}, \cite{Lo93}, \cite{Me:Co:Su}, 
\cite{Suaz:Sus}, \cite{Wolf81} and \cite{Yeon:Lee:Um:George:Pandey93} for a
general approach and known elementary solutions; a case related to Airy
functions is discussed in \cite{Lan:Sus} and our note deals with another
special case of transcendental solutions).\medskip

The corresponding Green function, or Feynman's propagator, can be found as
follows \cite{Cor-Sot:Lop:Sua:Sus}, \cite{Suaz:Sus}:%
\begin{equation}
\psi =G\left( x,y,t\right) =\frac{1}{\sqrt{2\pi i\mu _{0}\left( t\right) }}\
e^{i\left( \alpha _{0}\left( t\right) x^{2}+\beta _{0}\left( t\right)
xy+\gamma _{0}\left( t\right) y^{2}\right) },  \label{in2}
\end{equation}%
where%
\begin{eqnarray}
&&\alpha _{0}\left( t\right) =\frac{1}{4a\left( t\right) }\frac{\mu
_{0}^{\prime }\left( t\right) }{\mu _{0}\left( t\right) }-\frac{c\left(
t\right) }{2a\left( t\right) },  \label{in3} \\
&&\beta _{0}\left( t\right) =-\frac{\lambda \left( t\right) }{\mu _{0}\left(
t\right) },\qquad \lambda \left( t\right) =\exp \left( \int_{0}^{t}\left(
c\left( s\right) -d\left( s\right) \right) \ ds\right) ,  \label{in4} \\
&&\gamma _{0}\left( t\right) =\frac{1}{2\mu _{1}\left( 0\right) }\frac{\mu
_{1}\left( t\right) }{\mu _{0}\left( t\right) }+\frac{c\left( 0\right) }{%
2a\left( 0\right) }  \label{in5}
\end{eqnarray}%
and $\mu _{0}\left( t\right) $ and $\mu _{1}\left( t\right) $ are two
linearly independent solutions of the\ characteristic equation%
\begin{equation}
\mu ^{\prime \prime }-\tau \left( t\right) \mu ^{\prime }+4\sigma \left(
t\right) \mu =0  \label{in6}
\end{equation}%
with%
\begin{equation}
\tau \left( t\right) =\frac{a^{\prime }}{a}+2c-2d,\qquad \sigma \left(
t\right) =ab-cd+\frac{c}{2}\left( \frac{a^{\prime }}{a}-\frac{c^{\prime }}{c}%
\right)  \label{in7}
\end{equation}%
subject to initial conditions%
\begin{eqnarray}
&&\mu _{0}\left( 0\right) =0,\qquad \mu _{0}^{\prime }\left( 0\right)
=2a\left( 0\right) \neq 0,  \label{in8} \\
&&\mu _{1}\left( 0\right) \neq 0,\qquad \mu _{1}^{\prime }\left( 0\right) =0.
\label{in9}
\end{eqnarray}%
In this note, we present the time-dependent coefficient $\gamma _{0}\left(
t\right) $ in terms of two standard solutions $\mu _{0}$ and $\mu _{1}$ of
the characteristic equation (\ref{in6})--(\ref{in7}) with initial data (\ref%
{in8})--(\ref{in9}), respectively, instead of a previously used formula in
terms of only one solution $\mu _{0}$ (see, for example, \cite%
{Cor-Sot:Lop:Sua:Sus}, \cite{Cor-Sot:Sua:SusInv} and \cite{Suslov10}). Our
expression (\ref{in5}) can be verified by a direct differentiation:%
\begin{equation}
\left( \frac{\mu _{1}}{\mu _{0}}\right) ^{\prime }=\frac{\mu _{1}^{\prime
}\mu _{0}-\mu _{1}\mu _{0}^{\prime }}{\mu _{0}^{2}}=\frac{W\left( \mu
_{0},\mu _{1}\right) }{\mu _{0}^{2}},  \label{in10}
\end{equation}%
where the Wronskian can be found with the help of Abel's theorem as follows%
\begin{equation}
W\left( \mu _{0},\mu _{1}\right) =\text{constant\ }\lambda ^{2}\left(
t\right) a\left( t\right) .  \label{in11}
\end{equation}%
Then, as required,%
\begin{equation}
\frac{d\gamma _{0}}{dt}+a\left( t\right) \beta _{0}^{2}\left( t\right) =0
\label{in12}
\end{equation}%
and the constant term in (\ref{in5}) gives a correct asymptotic as $%
t\rightarrow 0^{+}$ (cf.~Refs.~\cite{Cor-Sot:Sus} and \cite{Suaz:Sus}). The
Green function (\ref{in2}) is an eigenfunction of the linear dynamical
invariant of Dodonov, Malkin, Man'ko and Trifonov \cite{Dod:Mal:Man75}, \cite%
{Dodonov:Man'koFIAN87}, \cite{Malkin:Man'ko79} and \cite{Malk:Man:Trif73}
(see also Theorem~1 of \cite{Suslov10}).\medskip

By the superposition principle the solution of the Cauchy initial value
problem can be presented in an integral form%
\begin{equation}
\psi \left( x,t\right) =\int_{-\infty }^{\infty }G\left( x,y,t\right) \ \chi
\left( y\right) \ dy,\quad \lim_{t\rightarrow 0^{+}}\psi \left( x,t\right)
=\chi \left( x\right)  \label{CauchyInVProb}
\end{equation}%
for a suitable initial function $\chi $ on $%
%TCIMACRO{\U{211d} }%
%BeginExpansion
\mathbb{R}
%EndExpansion
$ (a rigorous proof is given in \cite{Suaz:Sus} and uniqueness is analyzed
in \cite{Cor-Sot:Sua:SusInv}; another form of solution is provided by an
eigenfunction expansion \cite{Suslov10}; generalized coherent states and
transition amplitudes are discussed in a forthcoming paper \cite{Suslov10a}%
). In the next sections, we apply these general results to the case of the
degenerate parametric oscillator (\ref{at3}).

\section{Green's Function and Ince Equation}

For the Hamiltonian (\ref{at3}), variable coefficients are%
\begin{eqnarray}
&&a=\frac{\hslash }{2m}\left( 1+\frac{\lambda }{\omega }\cos \left( 2\omega
t\right) \right) ,  \label{ATa} \\
&&b=\frac{m\omega ^{2}}{2\hslash }\left( 1-\frac{\lambda }{\omega }\cos
\left( 2\omega t\right) \right) ,  \label{ATb} \\
&&c=d=\frac{\lambda }{2}\sin \left( 2\omega t\right)  \label{ATcd}
\end{eqnarray}%
and general expressions for Green's function (\ref{in2})--(\ref{in5}) can be
simplified by letting $\lambda \left( t\right) \equiv 1$ and $c\left(
0\right) =0.$ The characteristic equation is given by%
\begin{equation}
\mu ^{\prime \prime }+\frac{2\lambda \omega \sin \left( 2\omega t\right) }{%
\omega +\lambda \cos \left( 2\omega t\right) }\mu ^{\prime }+\frac{\omega
\left( \omega ^{2}-3\lambda ^{2}\right) -\lambda \left( \omega ^{2}+\lambda
^{2}\right) \cos \left( 2\omega t\right) }{\omega +\lambda \cos \left(
2\omega t\right) }\mu =0,  \label{InceEqCh}
\end{equation}%
which can be identified as a special case of the Ince equation \cite{Mag:Win}%
, \cite{Menn68}:%
\begin{equation}
\left( 1+a_{0}\cos 2s\right) y^{\prime \prime }\left( s\right) +b_{0}\sin
2s\ y^{\prime }\left( s\right) +\left( c_{0}+d_{0}\cos 2s\right) y\left(
s\right) =0,  \label{InceEq}
\end{equation}%
when $s=\omega t$ and parameters are given by%
\begin{equation}
a_{0}=\frac{\lambda }{\omega },\quad b_{0}=2a_{0}=2\frac{\lambda }{\omega }%
,\quad c_{0}=1-3\frac{\lambda ^{2}}{\omega ^{2}},\quad d_{0}=-\frac{\lambda 
}{\omega }\left( 1+\frac{\lambda ^{2}}{\omega ^{2}}\right) .
\label{InceParsCD}
\end{equation}

Traditionally, a special question which arises in the theory of Ince's
equation is the problem of the existence of periodic solutions. By
Theorem~7.1 on p.~93 of \cite{Mag:Win}, if Ince's equation (\ref{InceEq})
has two linearly independent solutions of period $\pi ,$ then the polynomial%
\begin{equation}
P\left( \xi \right) =2a_{0}\xi ^{2}-b_{0}\xi -d_{0}/2  \label{PolQ}
\end{equation}%
has a zero at one of the points $\xi =0,$ $\pm 1,$ $\pm 2,$ $...\ .$ If (\ref%
{InceEq}) has two linearly independent solutions of period $2\pi ,$ then%
\begin{equation}
Q\left( \xi \right) =2P\left( \xi -1/2\right)  \label{PlQSt}
\end{equation}%
vanishes for one of the values $\xi =0,$ $\pm 1,$ $\pm 2,$ $...\ .$\medskip

In the case of the degenerate parametric oscillator Hamiltonian (\ref%
{InceParsCD}), both of these polynomials are strictly positive:%
\begin{equation}
P\left( \xi \right) =\frac{2\lambda }{\omega }\left( \left( \xi -\frac{1}{2}%
\right) ^{2}+\frac{\lambda ^{2}}{4\omega ^{2}}\right) >0  \label{PolQin}
\end{equation}%
and%
\begin{equation}
Q\left( \xi \right) =\frac{4\lambda }{\omega }\left( \left( \xi -1\right)
^{2}+\frac{\lambda ^{2}}{4\omega ^{2}}\right) >0.  \label{PolQinSt}
\end{equation}%
Thus the corresponding Ince's equation (\ref{InceEqCh}) may not have two
periodic solutions of periods $\pi /\omega $ or $2\pi /\omega $ and our
standard solutions $\mu _{0}$ and $\mu _{1}$ that satisfy initial conditions
(\ref{in8})--(\ref{in9}) cannot be constructed in terms of simple variants
of the Fourier sine and cosine series:%
\begin{equation}
\mu _{1}\left( t\right) =\sum_{n=0}^{\infty }A_{2n}\cos \left( 2\omega
nt\right) ,\qquad \mu _{0}\left( t\right) =\sum_{n=1}^{\infty }B_{2n}\sin
\left( 2\omega nt\right) ,  \label{StSolI}
\end{equation}%
or%
\begin{equation}
\mu _{1}\left( t\right) =\sum_{n=0}^{\infty }A_{2n+1}\cos \left( \omega
\left( 2n+1\right) t\right) ,\qquad \mu _{0}\left( t\right)
=\sum_{n=0}^{\infty }B_{2n+1}\sin \left( \omega \left( 2n+1\right) t\right) ,
\label{StSolII}
\end{equation}%
when coefficients $A_{m}$ and $B_{m}$ satisfy certain three-term recurrence
relations (more details can be found in \cite{Mag:Win}). Therefore, the
degenerate parametric oscillator (\ref{at3}) motivates a detailed study of
non-periodic solutions of Ince's equation.

\section{The Dynamical Invariant and Time-Dependent Wave Functions}

Exact time-dependent wave functions of generalized harmonic oscillators (\ref%
{Schroedinger})--(\ref{GenHam}) can also be obtained as eigenfunctions of
the quadratic invariant in terms Hermite polynomials and certain solutions
of Ermakov-type equation (see Ref.~\cite{Suslov10} for a modern introduction
and references therein). The dynamical invariant of the Hamiltonian (\ref%
{at3}) can be derived from a general form presented in \cite%
{Cor-Sot:Sua:SusInv}, \cite{Suslov10} as follows:%
\begin{equation}
E\left( t\right) =\left( \mu \ p+\frac{m}{\hslash }\frac{\lambda \sin \left(
2\omega t\right) \mu -\mu ^{\prime }}{1+\left( \lambda /\omega \right) \cos
\left( 2\omega t\right) }\ x\right) ^{2}+\frac{C_{0}}{\mu ^{2}}\ x^{2},\quad
p=-i\frac{\partial }{\partial x},  \label{QuadInv}
\end{equation}%
where $C_{0}>0$ is a constant and function $\mu \left( t\right) $ satisfies
the nonlinear auxiliary equation:%
\begin{eqnarray}
&&\mu ^{\prime \prime }+\frac{2\lambda \omega \sin \left( 2\omega t\right) }{%
\omega +\lambda \cos \left( 2\omega t\right) }\mu ^{\prime }+\frac{\omega
\left( \omega ^{2}-3\lambda ^{2}\right) -\lambda \left( \omega ^{2}+\lambda
^{2}\right) \cos \left( 2\omega t\right) }{\omega +\lambda \cos \left(
2\omega t\right) }\mu  \notag \\
&&\qquad =C_{0}\left( \frac{\hslash }{m}\right) ^{2}\frac{\left( 1+\left(
\lambda /\omega \right) \cos \left( 2\omega t\right) \right) ^{2}}{\mu ^{3}}.
\label{InceErmEq}
\end{eqnarray}%
A general solution of this nonlinear equation can be found by the following
`law of cosines':%
\begin{equation}
\mu ^{2}\left( t\right) =Au^{2}\left( t\right) +Bv^{2}\left( t\right)
+2Cu\left( t\right) v\left( t\right)  \label{AuxSol}
\end{equation}%
in terms of two linearly independent solutions $u$ and $v$ of the
homogeneous equation. The constant $C_{0}$ is related to the Wronskian of
two linearly independent solutions $u$ and $v:$ 
\begin{equation}
AB-C^{2}=C_{0}\frac{\left( 2a\right) ^{2}}{W^{2}\left( u,v\right) },\quad
W\left( u,v\right) =uv^{\prime }-u^{\prime }v  \label{AuxSolWronskian}
\end{equation}%
(more details can be found in \cite{Suslov10}; see also references therein
regarding this so-called Pinney's solution). Thus solution of the
corresponding initial value problem is given by%
\begin{equation}
\mu ^{2}\left( t\right) =\left( \frac{\mu ^{\prime }\left( 0\right) }{%
2a\left( 0\right) }\mu _{0}\left( t\right) +\frac{\mu \left( 0\right) }{\mu
_{1}\left( 0\right) }\mu _{1}\left( t\right) \right) ^{2}+\frac{C_{0}}{\mu
^{2}\left( 0\right) }\mu _{0}^{2}\left( t\right) ,\quad \mu \left( 0\right)
\neq 0  \label{AuxSolIvp}
\end{equation}%
in terms of standard solutions $\mu _{0}$ and $\mu _{1}$ corresponding to
initial data (\ref{in8})--(\ref{in9}) of the homogeneous equation (\ref%
{InceErmEq}).\medskip

Time-dependent wave functions can be presented in the form%
\begin{equation}
\psi _{n}\left( x,t\right) =e^{-i\left( n+1/2\right) \varphi \left( t\right)
}\ \Psi _{n}\left( x,t\right)   \label{WaveFuncs}
\end{equation}%
(see, for example, \cite{Suslov10}, \cite{Yeon:Lee:Um:George:Pandey93} and
references therein), where%
\begin{equation}
\frac{d\varphi }{dt}=\sqrt{C_{0}}\frac{\hslash }{m}\frac{1+\left( \lambda
/\omega \right) \cos \left( 2\omega t\right) }{\mu ^{2}},\quad \varphi
\left( 0\right) =0  \label{WavePhase}
\end{equation}%
and orthonormal time-dependent eigenfunctions $\Psi _{n}\left( x,t\right) $
of the quadratic invariant (\ref{QuadInv}) are expressed in terms of Hermite
polynomials \cite{Ni:Su:Uv}:%
\begin{eqnarray}
&&\Psi _{n}\left( x,t\right) =D_{n}\exp \left( ix^{2}\frac{m}{2\hslash }%
\frac{\left( \mu ^{\prime }/\mu \right) -\lambda \sin \left( 2\omega
t\right) }{1+\left( \lambda /\omega \right) \cos \left( 2\omega t\right) }%
\right)   \label{InceInvEigfs} \\
&&\qquad \qquad \times e^{-x^{2}\sqrt{C_{0}}/\left( 2\mu ^{2}\right)
}H_{n}\left( x\frac{C_{0}^{1/4}}{\mu }\right) ,\quad \left\vert
D_{n}\right\vert ^{2}=\frac{C_{0}^{1/4}}{\sqrt{\pi }2^{n}n!\mu }.  \notag
\end{eqnarray}%
Then%
\begin{equation}
i\frac{\partial \psi _{n}}{\partial t}=H\psi _{n},\qquad E\psi _{n}=2\sqrt{%
C_{0}}\left( n+\frac{1}{2}\right) \psi _{n}.
\label{Schroedingereigenfunctions}
\end{equation}%
If the orthonormal initial wave function is given by%
\begin{equation}
\psi _{n}\left( x,0\right) =\sqrt{\frac{\varepsilon }{\sqrt{\pi }2^{n}n!}}\
e^{\left( i\delta -\varepsilon ^{2}/2\right) x^{2}}H_{n}\left( \varepsilon
x\right) ,  \label{InitFunc}
\end{equation}%
where $\varepsilon $ and $\delta $ are constants, then initial conditions
for the auxiliary equation (\ref{InceErmEq}) are given by%
\begin{equation}
\mu \left( 0\right) =\frac{C_{0}^{1/4}}{\varepsilon },\qquad \mu ^{\prime
}\left( 0\right) =2C_{0}^{1/4}\frac{\hslash }{m}\left( 1+\frac{\lambda }{%
\omega }\right) \frac{\delta }{\varepsilon }  \label{InitConds}
\end{equation}%
(the integral $C_{0}>0$ can be chosen at reader's convenience). Then
integration of Ermakov's equation (\ref{InceErmEq}), say with the help of
Pinney's solution (\ref{AuxSolIvp}), determines complete dynamics of the
oscillator-type wave function through the explicit formula (\ref%
{InceInvEigfs}). Further relations of these particular solutions of the Schr%
\"{o}dinger equation with the Cauchy initial value problem are discussed in 
\cite{Suslov10} (see Theorem~2).

\section{The $SU\left( 1,1\right) $ Symmetry}

The Hamiltonian (\ref{at2}) can also be rewritten as follows%
\begin{equation}
H\left( t\right) =\hslash \omega J_{0}-\hslash \lambda \left( e^{2i\omega
t}J_{+}+e^{-2i\omega t}J_{-}\right) ,  \label{sym1}
\end{equation}%
where generators of a non-compact $SU\left( 1,1\right) $ group are given by%
\begin{equation}
J_{+}=\frac{1}{2}\left( a^{\dagger }\right) ^{2},\qquad J_{-}=\frac{1}{2}%
a^{2},\qquad J_{0}=\frac{1}{4}\left( a^{\dagger }a+a^{\dagger }a\right) .
\label{sym2}
\end{equation}%
Then a use can be made of the group properties of the corresponding discrete
positive series $\mathcal{D}_{+}^{j}$ for further investigation of the
degenerate parametric oscillator. This is a `standard procedure' --- more
details can be found in Refs.~\cite{Lo93}, \cite{Malkin:Man'ko79}, \cite%
{Me:Co:Su}, \cite{Ni:Su:Uv} and \cite{Smir:Shit} and/or elsewhere.

\section{A Special Case}

An integrable special case $\lambda =\omega =m=\hslash =1$ of the degenerate
parametric oscillator has been recently considered by Meiler, Cordero-Soto
and Suslov \cite{Me:Co:Su}, \cite{Cor-Sot:Sus} (this Hamiltonian is a
simplest time-dependent quadratic integral of motion for the linear harmonic
oscillator \cite{Cor-Sot:Sua:SusInv}). Here, the Ince equation (\ref%
{InceEqCh}) simplifies to%
\begin{equation}
\mu ^{\prime \prime }+2\tan t\ \mu ^{\prime }-2\mu =0.  \label{MeCo-SoSuEq}
\end{equation}%
It has two elementary non-periodic standard solutions:%
\begin{eqnarray}
&&\mu _{0}=\cos t\sinh t+\sin t\cosh t,\qquad \mu _{0}\left( 0\right)
=0,\quad \mu _{0}^{\prime }\left( 0\right) =2,  \label{SolZero} \\
&&\mu _{1}=\cos t\cosh t+\sin t\sinh t,\qquad \mu _{1}\left( 0\right)
=1,\quad \mu _{1}^{\prime }\left( 0\right) =0  \label{SolOne}
\end{eqnarray}%
and Green's function is given in terms of trigonometric and hyperbolic
functions as follows%
\begin{eqnarray}
G\left( x,y,t\right)  &=&\frac{1}{\sqrt{2\pi i\left( \cos t\sinh t+\sin
t\cosh t\right) }}  \label{modGreen} \\
&&\times \exp \left( \frac{\left( x^{2}-y^{2}\right) \sin t\sinh
t+2xy-\left( x^{2}+y^{2}\right) \cos t\cosh t}{2i\left( \cos t\sinh t+\sin
t\cosh t\right) }\right)   \notag
\end{eqnarray}%
as a simple consequence of expressions (\ref{in2})--(\ref{in5}). It is worth
noting that formula (\ref{modGreen}) has been obtained in \cite{Me:Co:Su} by
a totally different approach using the $SU\left( 1,1\right) $-symmetry of $n$%
-dimensional oscillator wave functions and properties of the
Meixner--Pollaczek polynomials. More details can be found in\ \cite{Me:Co:Su}%
, \cite{Cor-Sot:Sus}.

\section{An Extension}

The degenerate parametric amplification with time-dependent amplitude and
phase had been considered by Raiford \cite{Raiford74}. The corresponding
Hamiltonian, without damping and neglecting high-frequency terms, has the
form%
\begin{equation}
H\left( t\right) =\frac{\hslash \omega }{2}\left( \widehat{a}\ \widehat{a}%
^{\dagger }+\widehat{a}^{\dagger }\ \widehat{a}\right) -\frac{\hslash
\lambda \left( t\right) }{2}\left( e^{i\left( 2\omega t+\delta \left(
t\right) \right) }\ \widehat{a}^{2}+e^{-i\left( 2\omega t+\delta \left(
t\right) \right) }\left( \widehat{a}^{\dagger }\right) ^{2}\right) .
\label{RaiHam}
\end{equation}%
In this model, the phenomenological coupling parameter $\lambda \left(
t\right) ,$ which describes the strength of the interaction between the
quantized signal of frequency $\omega $ and the classical pump of frequency $%
2\omega ,$ and the pump phase $\delta \left( t\right) $ are in general
functions of time as indicated. It includes the special case of the pump and
signal being off-resonance by a given amount $\epsilon ,$ i.~e., the pump
frequency being $2\omega +\epsilon ,$ by letting $\delta \left( t\right)
=\epsilon t$ and $\lambda \left( t\right) =\lambda ,$ a constant. More
details, including Heisenberg's equation of motion for the annihilation and
creation operators for the signal photons, can be found in the original
paper \cite{Raiford74}.

In the coordinate representation, the Hamiltonian (\ref{RaiHam}) is
rewritten as%
\begin{eqnarray}
&&H=\frac{1}{2m}\left( 1+\frac{\lambda \left( t\right) }{\omega }\cos \left(
2\omega t+\delta \left( t\right) \right) \right) p^{2}\label{RaiHamPX} \\
&&\qquad +\frac{m\omega ^{2}}{2}\left( 1-\frac{\lambda \left( t\right) }{%
\omega }\cos \left( 2\omega t+\delta \left( t\right) \right) \right) x^{2} 
\notag \\
&&\qquad \quad +\frac{\lambda }{2}\sin \left( 2\omega t+\delta \left(
t\right) \right) \left( px+xp\right) ,\qquad p=\frac{\hslash }{i}\frac{%
\partial }{\partial x}  \notag
\end{eqnarray}%
and Green's function can be found by using the method described in
section~2. Further details are left to the reader.

\section{A Conclusion}

In this Letter, we have constructed Green's function of the degenerate
parametric oscillator (\ref{at3}) in terms of standard non-periodic
solutions of Ince's equation. To the best of our knowledge, this oscillator
was introduced for the first time by Takahasi \cite{Taka65} in order to
describe the process of \ degenerate parametric amplification in quantum
optics (see also \cite{Louisel64}, \cite{Orszag}, \cite{Per:Hra:Jur}, \cite%
{Raiford70}, \cite{Raiford74}, \cite{Stoler74}, \cite{Walls:Milburn} and
references therein). The corresponding Hamiltonian had also been considered
later by Angelow and Trifonov \cite{Angelow:Trifonov95}, \cite{Angelow98} in
order to describe the light propagation in a nonlinear anisotropic
waveguide. Our observation motivates further investigation of properties of
the degenerate parametric oscillator including a systematic study of
corresponding non-periodic solutions of Ince's equation that seems to be
missing in mathematical literature. Moreover, the Hamiltonian (\ref{RaiHam}%
), which was considered by Raiford \cite{Raiford74} in the case of the
degenerate parametric amplification with time-dependent amplitude and phase,
requires a generalization of Ince's equation.

\noindent \textbf{Acknowledgments.\/} We thank Professor Carlos Castillo-Ch%
\'{a}vez, Professor Vladimir~I.~Man'ko, Professor Svetlana Roudenko and
Professor Dimiter A. Trifonov for support, valuable discussions and
encouragement. The authors are indebted to Professor Victor V.~Dodonov for
kindly pointing out papers \cite{Taka65}, \cite{Raiford74} and \cite%
{Stoler74} on the degenerate parametric amplification to our attention ---
his valuable comments have essentially helped to improve the paper. This
work is done as a part of the summer 2010 program on analysis of
Mathematical and Theoretical Biology Institute (MTBI) and Mathematical,
Computational and Modeling Sciences Center (MCMSC) at Arizona State
University. The MTBI/SUMS Summer Research Program is supported by The
National Science Foundation (DMS-0502349), The National Security Agency
(DOD-H982300710096), The Sloan Foundation and Arizona State University. One
of the authors (RCS) is supported by the following National Science
Foundation programs: Louis Stokes Alliances for Minority Participation
(LSAMP): NSF Cooperative Agreement No. HRD-0602425 (WAESO LSAMP Phase IV);
Alliances for Graduate Education and the Professoriate (AGEP): NSF
Cooperative Agreement No. HRD-0450137 (MGE@MSA AGEP Phase II).

\end{document}